\documentclass[a4]{nrc1}

\usepackage[dvips]{graphicx}
\usepackage[figuresright]{rotating}

\usepackage{amssymb}
\usepackage{amsmath}

\usepackage{cite}

\begin{document}

\title{Systematic trends in electronic properties of alkali hydrides}

\author{Mireille Aymar}
  \address[LAC]{Laboratoire Aim\'e Cotton, CNRS, B\^at. 505, Univ Paris-Sud, 91405 Orsay Cedex, France}
\author{Johannes Deiglmayr}
  \present{Physikalisches Institut, Universit\"at Freiburg, Hermann-Herder-Strasse 3, 79104 Freiburg, Germany}
\author{Olivier Dulieu}
  \correspond{olivier.dulieu@lac.u-psud.fr}

\shortauthor{Aymar, Deiglmayr, and Dulieu} 

\maketitle

\begin{abstract}
Obtaining  ultracold samples of dipolar molecules is a current challenge which requires an accurate knowledge of their electronic properties to guide  the ongoing experiments. Alkali hydride molecules have permanent dipole significantly larger than those of mixed alkali species and, as pointed out by Taylor-Juarros {\it et al.} [Eur. Phys. J. D {\bf 31}, 213 (2004)] and by Juarros {\it et al.} [Phys. Rev. A {\bf 73}, 041403 (2006)], are thus good candidates for molecule formation. In this paper, using a standard quantum chemistry approach based on pseudopotentials for atomic core representation, large Gaussian basis sets, and effective  core polarization  potential, we systematically investigate the electronic properties of the alkali hydrides LiH to CsH, in order to discuss general trends of their behavior. We computed (for the first time for NaH, KH, RbH, and CsH) the variation of their static polarizability with the internuclear distance. Moreover, in addition to potential curves, we determine accurate values of permanent and transition dipole moments for ground and excited states depending on the internuclear distance. The behavior of electronic properties of all alkali hydrides is compared to each other, in the light of the numerous other data available in the literature. Finally, the influence of the quality of the representation of the hydrogen electronic affinity in the approach on the results is discussed.



\end{abstract}

\section{Introduction}

Alkali hydrides have continuously attracted the interest of researchers in various areas. Indeed, they are among the molecules with the simplest electronic structure, allowing very detailed comparisons between different theoretical models, and with available experimental results. The astrophysical and cosmological relevance of LiH is well known \cite{dalgarno1995,bellini1994,maoli1994}, as it is believed to be formed in the early universe by radiative association of H and Li \cite{dalgarno1995}. The determination of the abundances of chemical elements in stellar atmospheres and the modeling of the chemistry of these elements in the early universe requires a knowledge of the cross sections for various collisional processes such as inelastic collisions between hydrogen and alkali atoms (Li, Na) \cite{belyaev1999,belyaev2003}, radiative charge transfer between alkali atoms and H$^+$ \cite{dutta2001a}, dissociative recombination \cite{curik2007}, or neutralization collisions between alkali ions and H$^-$ \cite{croft1999,mendez1990}. Molecular reaction dynamics of alkali atoms and molecular hydrogen involve alkali hydrides as products and a good description of these systems is required to analyze the reaction mechanism \cite{lee1999,lhermite1990}. C\^ot\'e {\it et al.} \cite{cote2000,derevianko2001a} suggested that the formation of Bose-Einstein condensates of a dilute gas of hydrogen could be achieved by cooling hydrogen atoms in a buffer gas of alkali atoms, which critically depends on the related scattering lengths. Recently Juarros {\it et al.} have  predicted that formation of ultracold LiH or NaH molecules is possible using various  processes such as stimulated Raman photoassociation or stimulated one-photon photoassociation \cite{taylor-juarros2004,juarros2006,juarros2006a}. Due to its large permanent dipole moment, the LiH molecule \cite{bethlem2002}, and possibly the NaH molecule, may be considered as candidates for Stark deceleration. Also, following the ideas of ref.\cite{friedrich1999a} related to alignment and orientation of polar molecules under the influence of combined static and laser fields, alkali hydrides could be good candidates for such experiments. An accurate knowledge of the electronic properties of these molecules are prerequisite to the extension of these investigations to heavier alkali hydrides.

Following previous work on alkali dimers in our group, we performed systematic computations of the electronic properties of alkali hydrides from LiH to CsH (Section \ref{sec:calcul}) as well as the monocations formed from these molecules. New results for the dependence on internuclear distance and on the vibrational index of their dipole polarizabilities are presented in Section \ref{sec:pola}. We also compare our results with numerous previous theoretical data for potential curves and permanent and transition dipole moments, identifying general trends for these quantities (Section \ref{sec:dipole}), and allowing for an analysis of their accuracy. In particular, the issue of the inclusion of a correction for the hydrogen electron affinity is discussed. Most of the computed results of the present work are available as supplementary material provided by the editor. In the following we will use atomic units for distances ($a_0$=0.052917720859~nm) and energies ($2R_{\infty}=219474.63137$~cm$^{-1}$) except where otherwise stated.

\section{Computational details and potential curves for neutral and ionic systems}
\label{sec:calcul}

As in our previous studies on alkali dimers, the alkali atoms are described by $\ell$-dependent pseudopotentials for the ionic cores \cite{durand1974,durand1975} including effective core polarization terms \cite{muller1984,foucrault1992}, and by a large set of uncontracted Gaussian functions for the valence electron. For all atoms but Cs we used the Gaussian basis sets labeled as "B" in ref.\cite{aymar2005,aymar2006a}, while for Cs we use the basis set labeled as "B'" in ref.\cite{aymar2006}. We also use the same core-polarization potentials (CPP) than in our previous work, with the reported values for the cut-off radii. For the H atom we set up a new large $[10s5p2d]$ basis set of uncontracted Gaussian functions with the following exponents: 195.5, 27.6, 6.3, 1.8, 0.5983, 0.22, 0.09, 0.04, 0.019, 0.006 for $s$ basis functions; 1.05, 0.22, 0.0669, 0.024270, 0.00669 for $p$ basis functions; 0.34, 0.061 for $d$ basis functions. This choice globally improves the calculated atomic level energies compared to previous works. We obtain the following discrepancies: $\Delta E_{1s}=3.2$~cm$^{-1}$, $\Delta E_{2s}=2.1$~cm$^{-1}$, $\Delta E_{2p}=7.4$~cm$^{-1}$,  $\Delta E_{3s}=29$~cm$^{-1}$, $\Delta E_{3p}=51$~cm$^{-1}$. For instance, Huzinaga {\it et al.} \cite{huzinaga1965} used a large basis set of contracted Gaussian orbitals which gave exact energies for the $1s$, $2s$, $2p$, $3s$ and $3p$ levels. In their recent series of papers, Gad\'ea and coworkers \cite{boutalib1992,khelifi2002,khelifi2002a,zrafi2006} designed a $[7s3p2d]$ Gaussian basis set contracted to $[5s3p2d]$, close to the one of Geum {\it et al.} \cite{geum2001} who used a $[7s3p2d]$ contracted to $[6s3p2d]$ basis set. The former authors obtained deviations of $\Delta E_{1s} \sim  11$~cm$^{-1}$ and $\Delta E_{2s}\sim 700$~cm$^{-1}$ \cite{boutalib1992}, and the latter $\Delta E_{1s}\sim  64$cm$^{-1}$ and $\Delta E_{2s}\sim 35$~cm$^{-1}$. Moreover our basis also provides an improved hydrogen electronic affinity with a deviation of about 140~cm$^{-1}$ from the exact value of ref.\cite{pekeris1962}, i.e. three times smaller than the value quoted by Boutalib and Gad\'ea \cite{boutalib1992} but slightly larger than the value of Geum {\it et al.} ($\sim 111$~cm$^{-1}$).

The molecular calculations are performed just like in our previous papers \cite{aymar2005,aymar2006,aymar2007}. Molecular orbitals are determined by a restricted Hartree-Fock calculation including core polarization, yielding electronic energies and wave functions of the molecular ion. Electronic energies and wave functions of the neutral system are deduced from a full configuration interaction (CI) of two-electron configurations, through the CIPSI quantum chemistry code of the quantum chemistry group in Toulouse (France) \cite{huron1973}. Our procedure automatically provides potential curves for the related alkali hydride ions treated as effective one-electron systems, and it is worthwhile to examine them as a first check of its accuracy compared to other approaches. It is striking to see in Table \ref{tab:ions} that even for the ground state of such simple systems, published values vary significantly for the binding energy of LiH$^+$ and NaH$^+$, while only few data are available up to now for the heaviest species KH$^+$, RbH$^+$, and CsH$^+$. To our knowledge only one experimental value has been published up to now, namely the CsH$^+$ binding energy extracted from differential cross section measurements \cite{scheidt1978}. Very few measurements have been reported for other molecular states. We also list in Table \ref{tab:ions} data for the first excited state $A^2\Sigma^+$ which has been widely studied in the context of the theoretical calculation of cross sections for charge exchange reactions \cite{kubach1981,kimura1982a,allan1986,dalgarno1995,dutta2001a,watanabe2002}.

Most of the calculations of alkali hydride ion potential curves have been carried out with effective core potentials. In a series of recent papers, Magnier \cite{magnier2004a,magnier2005,magnier2006} extended the work of Alikacem and Aubert-Fr\'econ \cite{alikacem1985} by including effective core polarization potentials to their Klapisch  model potential approach  \cite{klapisch1969}. Various kinds of pseudopotentials have been used by several authors  \cite{valance1978,melius1979,fuentealba1982,szentpaly1982,kubach1981,kimura1982a,watanabe2002}, the closest approach to ours being the work by Berriche and Gad\'ea \cite{berriche1995} for LiH$^+$. A few {\it ab initio} all-electron calculations have also been performed many years ago, by Rosmus and Meyer \cite{rosmus1977} with the coupled electron pair approach, by Vojtik \cite{vojtik1990} using the MRDCI  approach, and by Dalgarno {\it et al.} \cite{dalgarno1995} and Olson {\it et al.} \cite{olson1980} with SCF and CI approaches, for the treatment of radiative association in an astrophysical environment. It is tedious to achieve an enlightening comparison of our data with these works, as most of them are quite old and were probably limited by computational facilities at that time. Moreover, several calculations reported binding energies $D_e$ for the $A$ excited state only, with which our results reasonably agree, but without reporting similar data for the ground state. Our values for the LiH$^+$ ground state agree well with those of Magnier \cite{magnier2004a}, but are slightly larger than those obtained by the same author on  NaH$^+$ and KH$^+$ \cite{magnier2005,magnier2006}. For the $A$ state the present $D_e$ values for LiH$^+$ and NaH$^+$ are almost identical to those predicted by Magnier \cite{magnier2004a,magnier2005} but slightly smaller for KH$^+$ \cite{magnier2006}. Also, the depth of the potential well of the $A$ state of CsH$^+$ lies within the error of the data reported in ref.\cite{scheidt1978}.

\begin{table}[h]
\begin{tabular} {|c|c|c|c||c|c|}
\hline
\multicolumn{2}{|c|}{ }&\multicolumn{2}{c||}{$X^2 \Sigma^+$} &\multicolumn{2}{c|}{$A^2 \Sigma^+$}\\ \hline
Ion & Reference                                          &$R_e$&$D_e$(eV)&$R_e$& $D_e$(eV) \\ \hline
LiH$^+$&This work                                        &4.11 &0.1426&7.44&0.4994\\
       &Rosmus and Meyer \cite{rosmus1977}               &4.12 &0.13&&\\
       &Fuentealba {\it et al.} \cite{fuentealba1982}     &4.12 &0.14&&\\
       &Alikacem and Aubert Fr\'econ \cite{alikacem1985}&4.08 &0.13  &7.35&0.50\\
       &Vojtik {\it et al.} \cite{vojtik1990}            &4.15 &0.13  &7.27&0.41\\
       &Berriche {\it et al.} \cite{berriche1995}         &4.11 &0.141 &7.45&0.497\\
       &Dalgarno {\it et al.} \cite{dalgarno1995}         &4.13 &0.1404&7.39&0.48\\
       &Kimura {\it et al.} \cite{kimura1982a}             &     &      &7.47&0.49\\
       &Magnier \cite{magnier2004a}                      &4.15 &0.13  &7.46&0.495\\ \hline
NaH$^+$&This work                                        &4.85 &0.083  &7.88&0.4594\\
       &Valance \cite{valance1978}                       &5.8  &0.02  &8.2 &0.34\\
       &Rosmus and Meyer \cite{rosmus1977}               & 4.90& 0.14&&\\
       &Melius {\it et al.} \cite{melius1979}             & 5.1 &0.061&&\\
       &Olson {\it et al.} \cite{olson1980}               & 5.1 & 0.061&7.98&0.469\\
       &Liu {\it et al.} \cite{liu1981}                   & 5.1 &0.064&&\\
       &Fuentealba {\it et al.} \cite{fuentealba1982}     &4.86 &0.08 &&\\
       &Watanabe {\it et al.} \cite{watanabe2002}         & 4.65&0.103 &7.8&0.435\\
       &Kubach and Sidis \cite{kubach1981}               &     &      &8.7&0.39\\
       &Kimura {\it et al.} \cite{kimura1982a}             &     &      &7.89&0.46\\
       &Allan \cite{allan1986}                           &     &      &7.88&0.46\\
       &Magnier \cite{magnier2005}                       & 4.9 &0.0615&7.86&0.4602\\ \hline
KH $^+$&This work                                        &5.66 &0.062 &8.68&0.6481\\
       &Valance \cite{valance1978}                      & 10. & 0.025& 8.6&0.49\\
       &Fuentealba {\it et al.} \cite{fuentealba1982}     &5.88 &0.05&&\\
       &Olson {\it et al.} \cite{olson1980}               &5.66 &0.054 &8.74&0.614\\
       &Melius {\it et al.} \cite{melius1979}             & 6.5 &0.022&&\\
       &Kubach and Sidis \cite{kubach1981}               &     &      &9.4 &0.49\\
       &Kimura {\it et al.} \cite{kimura1982a}             &     &      &8.68&0.68\\
       &Watanabe {\it et al.} \cite{watanabe2002}         &5.   &0.136 &8.45&0.625\\
       &Magnier \cite{magnier2006}                       & 5.52&0.0476&8.58&0.6607\\ \hline
RbH$^+$&This work                                        &5.82&0.6824&8.58&0.6824\\
       &Valance \cite{valance1978}                      &13.4&0.012 &8.8&0.50\\
       &Von Szenptaly {\it et al.}  \cite{szentpaly1982}  &5.83&0.06&&\\
       &Kubach and Sidis \cite{kubach1981}               &    &      &10.4&0.55\\
       &Kimura {\it et al.} \cite{kimura1982a}             &    &      &8.87&0.73\\ \hline
CsH$^+$&This work                                        &5.98&0.0809&9.36&0.793\\
       &Valance \cite{valance1978}                      &17.6&0.03  &10.4&0.68\\
       &Von Szenptaly {\it et al.}  \cite{szentpaly1982}  &6.14&0.03  &    &\\
       &Scheidt {\it et al.} \cite{scheidt1978}            &    &      &10$\pm 1$ &0.77$\pm$ 0.05\\
       &Kubach and Sidis \cite{kubach1981}               &    &      &10.4&0.71\\
       &Kimura {\it et al.} \cite{kimura1982a}             &    &      &9.20&0.86\\ \hline
\end{tabular}
\caption {Equilibrium distances $R_e$ (in a.u.) and binding energies $D_e$ for the $X^2 \Sigma^+$ and $A^2\Sigma^+$ states of alkali hydride cations obtained in the present work, compared to previous predictions.}
\label{tab:ions}
\end {table}

Next, the accuracy of our full CI results yielded by our treatment of the neutral molecule as an effective two-electron system, can be analyzed from Tables \ref{tab:bond} and \ref{tab:spectro}, which display the equilibrium distances $R_e$, the harmonic constant $\omega_e$ and the potential well depth of the alkali hydrides electronic states together with available experimental values and with several other recent theoretical results. We concentrate our analysis on the fourth lowest $^1\Sigma^+$ and two lowest $^1\Pi$ electronic states, as they can be reached by electric dipole transition from the ground state. However, we also performed calculations for triplet states, and all these results are collected in the supplementary material attached to this paper. We label the molecular states according to the standard spectroscopic notation, i.e. $X, A, C, E$ corresponding to the four lowest $^1\Sigma^+$ states, and $B \equiv (1)^1\Pi$. Similarly, triplet states are labeled with the lowercase indexes $a$, $b$, and $c$, for the $(1)^3\Sigma^+$, $(1)^3\Pi$, and $(2)^3\Sigma^+$, respectively. In refs.\cite{boutalib1992,khelifi2002,khelifi2002a,zrafi2006}, the authors labeled the (4)$^1\Sigma^+$ as $D$ state. Our convention for the sign of the permanent dipole is that a positive value corresponds to the charge distribution A$^+$ H$^-$, where A is an alkali atom.

The extensive review by Stwalley {\it et al.} \cite{stwalley1991} addresses most of the works already published on alkali hydrides. In particular, this review reports on experimental studies of the first $^1\Sigma^+$ state (the $A$ state) from NaH to CsH. Vidal {\it et al.} \cite{vidal1982,vidal1984} investigated the spectroscopy of $A$  and $B$ (the lowest $^1\Pi$) states, and Yang {\it et al.} \cite{yang1980} and Hsieh {\it et al.} \cite{hsieh1978} the $A$ state in KH and in CsH, respectively.

We briefly recall below some of the theoretical calculations, mainly those yielding tables of permanent or transition dipole moments, for further reference in the next sections. {\it Ab initio} multiconfiguration self-consistent-field calculations with configuration interaction (MCSCF-CI) of permanent and transition dipole moments of LiH have been reported by Docken and Hinze \cite{docken1972} using a Slater-type basis set, in contrast with Gaussian basis sets employed in the present calculations. Partridge and Langhoff \cite{partridge1981,partridge1981a} repeated  such computations with an extended Slater-type basis set. Roos and Sadlej \cite{roos1982} used the complete active space SCF (CASSCF) approach to calculate the permanent dipole moment and the polarizabilities of the ground state as functions of the interatomic distance. Sachs {\it et al.} determined the electronic properties of  several states of NaH \cite{sachs1975,sachs1975a}. Langhoff {\it et al.} extended their previous investigations to NaH, KH, and RbH \cite{langhoff1986} using near Hartree-Fock quality Slater basis sets and incorporating electron correlation through a coupled-pair formalism. Laskowski {\it et al.} \cite{laskowski1981} studied the two lowest $^1\Sigma^+$ states in CsH, using a similar approach to ours: effective $\ell$-dependent pseudopotentials derived from relativistic Hartree-Fock calculations and including core polarization terms for atomic core representation, and Gaussian basis sets in the CI calculation. Carnell {\it et al.} \cite{carnell1989} investigated the first seventeen states of CsH with {\it ab initio} multi-reference configuration interaction (MRDCI) calculations. Combining existing experimental data and ab-initio calculations, Zemke {\it et al.} determined the dipole moment for the $A^1\Sigma^+\rightarrow X^1\Sigma^+$, $B^1\Pi \rightarrow X^1\Sigma^+$, and $B^1\Pi \rightarrow A^1\Sigma^+$ transitions and related radiative transition probabilities in LiH \cite{zemke1978a,zemke1978b}, and the permanent dipole moment for the $X$ and $A$ states in NaH \cite{zemke1984}. Finally, Camacho {\it et al.} \cite{camacho1998} built the $A^1\Sigma^+\rightarrow X^1\Sigma^+$ transition dipole moment of KH from available experimental data.

More recently Gad\'ea and coworkers have performed a series of studies of the adiabatic and diabatic molecular states of alkali hydrides \cite{boutalib1992,khelifi2002,khelifi2002a,zrafi2006}. As in the present work, the authors used the $\ell$-dependent pseudopotentials of Durand and Barthelat \cite{durand1974,durand1975}, and core-valence correlation from ref.\cite{muller1984,foucrault1992}. Furthermore, an $R$-dependent correction of the ion-pair diabatic curves is introduced to account for basis set limitations and to ensure an improved value for the ground state energy. Permanent dipole moments for several $^{1,3}\Sigma^+$  states and transition dipole moments between $^{1}\Sigma^+$ states have been displayed for KH \cite{khelifi2002}, RbH \cite{khelifi2002a}, and CsH \cite{zrafi2006}. Khelifi {\it et al.} have determined the radiative and nonradiative lifetimes of the $A^1\Sigma^+$ and $C^1\Sigma^+$ vibrational levels of the KH molecule \cite{khelifi2006,khelifi2007}.

In Table \ref{tab:bond}, we see that theoretical values for $R_e$ are systematically slightly smaller than the experimental ones, especially for CsH. This could be due to the absence of short-range repulsion terms in the ECP calculations. Our resulting values for $D_e$ are slightly larger than similar calculations of Gad\'ea and coworkers (column (a) in the Table) \cite{boutalib1992,khelifi2002,khelifi2002a,zrafi2006}, which is a manifestation of the influence of the large basis we used in our work. However it is well known that ECP-type calculation is not a perfect variational approach, so that it may happen that the computed well-depth exceeds the experimental one, especially when large ionic cores are involved (see the CsH case). This pattern is also visible in results from Dolg \cite{dolg1996}. They rely on fully relativistic pseudopotentials, and molecular calculations are based on Dirac-Hartree-Fock and CI approaches. Core-polarization terms have been introduced, as well as corrections for core-core repulsion. Equilibrium distances are also found shorter than the experimental ones, while $D_e$ are larger than the experimental ones for all species. When an empirical correction for the hydrogen electronic affinity is included by Gad\'ea and coworkers for KH, RbH, and CsH (see section \ref{sec:dipole}), the values for $D_e$ are increased and become closer to the experimental ones. However, such a correction cannot be considered as a variational procedure, so that their $D_e$ value for CsH exceeds the experimental one by about 462~cm$^{-1}$. We will comment more about the effect of their ion-pair correction on dipole moments in section \ref{sec:dipole}.

\begin{table}[h]
\center
\begin{tabular} {|cc|ccc|cccc|}
 \hline
  &     &\multicolumn{3}{|c|}{This work}&\multicolumn{4}{c|}{other works}\\
  &state&$R_e$& $D_e$&$\omega_e$&ref.&$R_e$& $D_e$&$\omega_e$ \\ \hline
LiH&$X$& 3.002&20166.8&1398&\cite{crawford1935}$^+$&3.015&20287.7 $\pm0.3$&1406.9 \\
   &   &      &       &    &\cite{boutalib1992}    &3.007&20174&\\
   &   &      &       &    &\cite{dolg1996}        &3.0  &20123&1391  \\ \hline
NaH&$X$&3.54  &15671.3&1163&\cite{stwalley1991}$^+$&3.566&15900$\pm 100$&1171.5\\
   &   &      &       &    &\cite{dolg1996}        &3.254&16050&1163  \\ \hline
KH &$X$&4.168 &14500.0&952 &\cite{stwalley1991}$^+$&4.23&14772.7$\pm 0.6$&986.6\\
   &   &      &       &    &\cite{khelifi2002}$^a$ &4.19&14365& \\
   &   &      &       &    &\cite{khelifi2002}$^b$ &4.19&14750.4&\\
   &   &      &       &    &\cite{dolg1996}        &4.2&14937   &961\\ \hline
RbH&$X$&4.395 &14098.8&885 &\cite{stwalley1991}$^+$&4.47&14580$\pm 600$&937.1\\
   &   &      &       &    &\cite{khelifi2002a}$^a$&4.40&13940& \\
   &   &      &       &    &\cite{khelifi2002a}$^b$&4.40&14323.2& \\
   &   &      &       &    &\cite{dolg1996}        &4.367&14800&912  \\ \hline
CsH&$X$&4.464 &15019.3&802 &\cite{stwalley1991}$^+$&4.71&14791.2$\pm 2$& 891.2\\
   &   &      &       &    &\cite{zrafi2006}$^a$   &4.47& 14878& \\
   &   &      &       &    &\cite{zrafi2006}$^b$   &4.48&15253.4& \\
   &   &      &       &    &\cite{dolg1996}        &4.626&14897&881\\ \hline
\end{tabular}
\caption {Equilibrium distances $R_e$ (in a.u.), harmonic constants $\omega_e$ and potential well depths $D_e$ (in cm$^{-1}$) of the ground state of alkali hydrides, compared to available experimental values, and recent theoretical determinations. The (a) and (b) labels refer to the calculation without, and with the correction for hydrogen electronic affinity, respectively (see text). Note that ref.\cite{stwalley1991} is a review paper which collected results from many experiments, which explains the apparently huge variations of the error limits.}
\label{tab:bond}
\end {table}

We summarized the main spectroscopic constants of the four lowest $^1\Sigma^+$ excited states, and of the lowest $^1\Pi$ state in Table \ref{tab:spectro}, compared to experimental observations, and to selected theoretical predictions. As already pointed out, Gad\'ea and coworkers
\cite{boutalib1992,khelifi2002,khelifi2002a,zrafi2006} have investigated all systems but NaH and their papers include extensive comparisons with other predictions which are not duplicated here. For NaH our values are compared to those of Lee {\it et al} \cite{lee2000a}, who used the large-core pseudopotential of the Stuttgart  group \cite{leininger1996} and accounted for core polarization following ref.\cite{muller1984}. SCF and CI calculations have been performed with the MOLPRO package. The $X$ and $A$ states of KH and RbH have been investigated by Garcia {\it et al} \cite{garcia1998} using a small-core pseudopotential \cite{leininger1996}, and the difference-dedicated configuration interaction (DDCI) method to account for core-valence correlation effects. Our values are generally in good agreement with those of Gad\'ea and coworkers although the Gaussian basis sets used in those work for describing the alkali and hydrogen atoms  differ from ours. It is difficult to know if the somewhat large differences with the values for NaH of
\cite{lee2000a} are related to the differences in the pseudopotential or to those in the basis sets.

\begin{table}[t]
\center
\begin{tabular} {|cc|ccc|cccc|}
 \hline
  &     &\multicolumn{3}{|c|}{This work}&\multicolumn{4}{c|}{other works}\\
  &state&$R_e$& $D_e$&$\omega_e$&ref.&$R_e$& $D_e$&$\omega_e$ \\ \hline
LiH&A&4.82& 8698&241&\cite{boutalib1992}&4.87&8689&\\
   & &    &     & &\cite{vidal1982}&4.91&8685.6 $\pm$0.3&244 \cite{crawford1935}\\
   &C$^1$&3.83&1267&390&\cite{boutalib1992}&3.825&1276.8& \\
   &C$^2$&10.15&8361&390&\cite{boutalib1992}&10.21&8444& \\
   &E&5.31&3727&516& \cite{boutalib1992}& 5.358&2839& \\
  &(5)&4.1 &1455&486& \cite{boutalib1992}& 4.094&1574& \\
   &B&4.52&251&243& \cite{vidal1984}& 4.5&288.9$\pm0.2$& \\ \hline
NaH&A&5.97&10045&318& \cite{lee2000a}& 6.01&9997& \\
   & &    &     & &\cite{sachs1975}&6.186&9701&333.1 \cite{sachs1975b}\\
   & &    &     & &\cite{stwalley1991}$^+$& 6.03&10143&317.56 \\
   &C$^1$& 4.45&661&432 & \cite{lee2000a}&  4.48&612&\\
   &C$^2$&11.75&6501&219& &   & &\\
   &E&5.23&3709&683& \cite{lee2000a} & 5.24&4738& \\
   &(5)& 4.57&1121&432&\cite{lee2000a} & 4.63&3927& \\
   &B& 5.1&282&214&                 &     &    &\\ \hline
KH &A& 6.95&8884&243&\cite{khelifi2002a}&7.05&8946.3&  \\
   & &     &    & &\cite{jeung1983a}&7.18&8710&288\\
   & &     &    & &\cite{garcia1998}&6.944&8590&245.9\\
   & &     &    & &\cite{yang1980}$^+$&7.11&8698&222.74$\pm 0.16$ \\
   &C& 13.3&6471&163&\cite{khelifi2002a}& 13.63&6584.5& \\
   &E& 5.59&970.&232&\cite{khelifi2002a}& 5.65&873.5.& \\
   &(5)& 5.27&1028&362&\cite{khelifi2002a}&5.28&1026.9& \\
   &B& 5.84&294&186&\cite{khelifi2002a}& 5.34&801& \\ \hline
RbH&A& 7.17&9053&217&\cite{khelifi2002}& 7.3&8711.2&   \\
   & &     &    & &\cite{garcia1998}&6.92&8710&232\\
   & &     &    & &\cite{stwalley1991}$^+$&6.270&8941&211.74 \\
   &C$^1$&5.61&3568&215&\cite{khelifi2002}&  5.65&3545.9& \\
   &C$^2$&14.1&5387&197&\cite{khelifi2002}&14.4&5509& \\
   &E& 5.74&553&245& \cite{khelifi2002}& 5.77&504.3& \\
   &(5)& 5.5&1162&353&\cite{khelifi2002}& 5.6&1133.3& \\
   &B&5.89&355&213&\cite{khelifi2002}& 5.92&377.3& \\ \hline
CsH&A&7.31&8825&192&\cite{zrafi2006}& 7.55&8851.9&  \\
   & &    &    & &\cite{laskowski1981}&7.52&7500&\\
   & &    &    & &\cite{jeung1983b}&8.10&8630 &196 \\
   & &    &    & &\cite{carnell1989}&7.46&7767&196\\
   & &    &    & &\cite{stwalley1991}$^+$& 7.52&8130& 196\\
   &C$^1$&5.63&2543&301&\cite{zrafi2006} & 5.63&2462.5& \\
   &C$^2$&14.8&2175&111&\cite{zrafi2006} &29&3785& \\
   & &    &    & &\cite{jeung1983b}& 6.&2180&242\\
   & &    &    & &\cite{carnell1989}&  &1930&144\\
   &E&5.83&659&264&\cite{zrafi2006} & 5.86&642.4&  \\
   &(5)& 5.69&1288&332&\cite{zrafi2006}&5.72&1354.36&  \\
   &B& 6.01&576&228&\cite{zrafi2006}&  6.04&551.89&  \\ \hline
\end{tabular}
\caption {Main spectroscopic constants of excited singlet electronic states of alkali hydrides, for the four lowest states of $^1\Sigma^+$ symmetry ($A$, $C$, $E$ and (5)) and the lowest state ($B$) of $^1\Pi$ symmetry. The exponents 1 and 2 labeling the $C$ state refers to the inner and the outer potential well, respectively. Equilibrium distances $R_e$ are in bohr radii, depth of potential wells $D_e$ and harmonic constants $\omega_e$ in cm$^{-1}$. Comparison is provided with available experimental results, and with a selection of recently published calculations (mainly from the series of papers of Gad\'ea and coworkers.}
\label{tab:spectro}
\end {table}

\section{Static dipole polarizabilities}
\label{sec:pola}

\begin{figure}
\includegraphics[width=0.8\columnwidth]{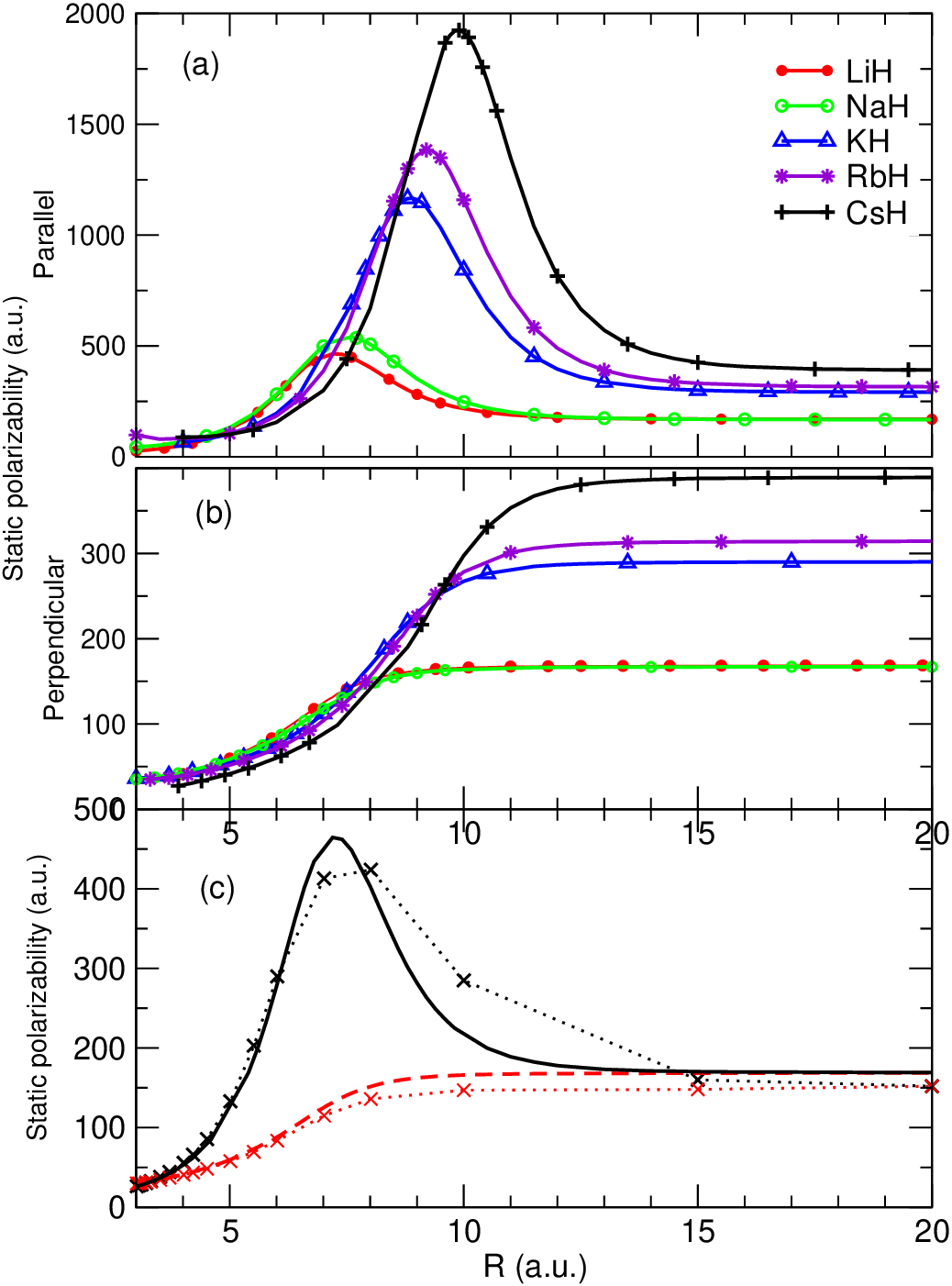}
\caption{\label{fig:pola} (a) $\alpha _{\parallel}$  and  (b) $\alpha_{\perp}$  components of the static dipole polarizability of the alkali hydride ground state  as functions of the internuclear distance. (c) For LiH, our results for $\alpha _{\parallel}$(full line) and $\alpha_{\perp}$ (dashed lines) are compared with those of Roos and Sadlej  \cite{roos1982} (X-es).}
\end{figure}

If the $z$ axis is chosen along the internuclear axis in the molecule-fixed reference frame ($x$,$y$,$z$), there are two independent components of the molecular polarizability tensor, i.e., the parallel component $\alpha_{\parallel} \equiv \alpha _{zz}$ and the perpendicular one $\alpha_{\perp} \equiv \alpha_{xx} = \alpha_{yy}$. Two related quantities are usually defined: the average polarizability  $\alpha=(\alpha _{\parallel} +2\alpha_{\perp})/3$ and the polarizability anisotropy $\gamma = \alpha_{\parallel}-\alpha_{\perp}$. As in our previous work on the static dipole polarizabilities of alkali dimers \cite{deiglmayr2008}, we determined the static dipole polarizabilities of alkali hydrides in their ground state with the finite-field method \cite{cohen1965}, using electric fields between 0.0003 and 0.0005~a.u. to remain in the perturbative regime. We calculated both components of the static dipole polarizability for all systems as functions of the internuclear distance $R$ (Figure \ref{fig:pola}). For all systems, the $R$-variations of $\alpha_{\parallel}$ and $\alpha_{\perp}$ are similar, with an increase of $\alpha_{\parallel}$ with increasing alkali mass, and a smooth increase of $\alpha_{\perp}$ with $R$. To our knowledge, no other $R$-variation of the alkali hydride dipole polarizability has been previously published, except for LiH in ref.\cite{roos1982} which displays a dependency in reasonable agreement with the present one (Figure \ref{fig:pola}(c)). Looking at large $R$, we note however that the lithium polarizability seems to be slightly underestimated in ref.\cite{roos1982}. Also, Kolos and Wolniewicz \cite{kolos1967} found similar variations of the H$_2$ polarizabilities than the present ones, but with smaller values, as expected.

\begin{figure}
\includegraphics[width=0.8\columnwidth]{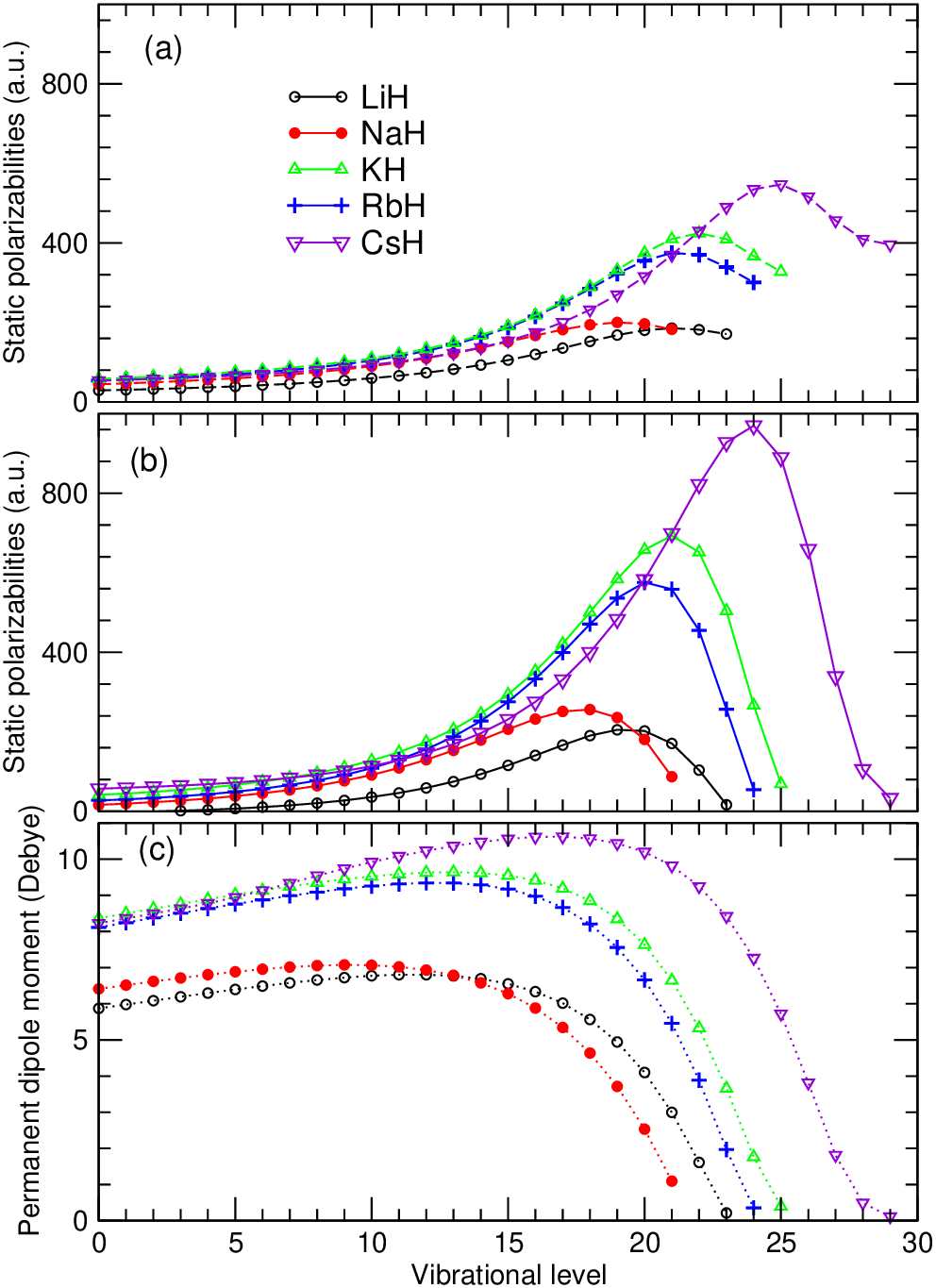}
\caption{\label{fig:pola-v} (a) average polarizability (dashed lines), (b) polarizability anisotropy (full lines) (in au.), and (b) permanent dipole moment (in Debye, with 1 a.u.=2.54158059~D) of the alkali hydride ground state, as functions of the vibrational level.}
\end{figure}

By integrating the polarizabilities over the vibrational wave functions, we deduce the $v$-dependency of $\alpha$ and of $\gamma$, which regularly increase from LiH to CsH, just like for the series of alkali pairs (Figure \ref{fig:pola-v}(a)). The anisotropies reach quite large values (up to 967 au. for CsH at $v=24$, i.e. two times larger than the cesium atomic polarizability), magnifying the influence of the hydrogen atom when the molecule is vibrationally excited, in contrast with the $v=0$ case. No other values have been published in the past. We also display in Figure \ref{fig:pola}(b) the permanent dipole moment of the alkali hydride ground state which show similar $v$-dependence for all the systems, with a maximum followed by a rapid decrease to zero, as expected from comparable calculations for alkali pairs \cite{aymar2005}.

In our previous work about the ground state polarizabilities of all homonuclear and heteronuclear alkali diatomics \cite{deiglmayr2008}, both $\alpha_{\parallel}$ and $\alpha_{\perp}$ components were found to vary linearly with an effective volume $V_{eff}=4\pi(R_e)^3/3$, $\alpha_{\parallel}$ varying two times faster than $\alpha_{\perp}$ with $V_{eff}$. This suggested the picture of an effective elliptic charge distribution for the alkali pair at the equilibrium distance. In alkali hydrides, this scaling law remains valid only for $\alpha_{\parallel}$, while $\alpha_{\perp}$ does not monotonically vary along the series of alkali hydrides. This may be due to the weakness of the hydrogen polarizability, which is at least one order of magnitude smaller (4.5 a.u. \cite{landau1967}) than the alkali ones (\cite{deiglmayr2008}). Therefore the effective elliptic charge distribution of the molecule is progressively dominated by the polarizability of the alkali atom, along the series from Li to Cs.

Most previous theoretical works reported polarizability values at the experimental equilibrium distance $R_e^{exp}$ of the ground state, which are assembled in Table \ref{tab:pola}, together with the present results for $\alpha_{\parallel}$, $\alpha_{\perp}$, $\alpha$ and $\gamma$ calculated either at $R_e^{exp}$ \cite{stwalley1991} or for $v=0$. The LiH molecule is the most extensively studied system, and we give only the most significant values. Additional references may be found in the quoted papers. Coupled Hartree-Fock perturbation theory has been used by Lazzeretti {\it et al.} \cite{lazzeretti1981}, and multiconfigurational time-dependent  Hartree-Fock (MCTDHF) method by Sasagane {\it et al.} \cite{sasagane1990}. Multiconfiguration self-consistent field (SCF) approaches have been used at various levels by several authors \cite{gready1977b,maroulis1986,roos1982,bishop1985}. In these papers, the finite field method \cite{cohen1965} or the charge perturbative approach \cite{maroulis1986} have been used to  extract polarizabilities. A time-dependent gauge invariant method (TDGI) has been used by R\'erat {\it et al.} \cite{rerat1992}, while Vrbik {\it et al.} \cite{vrbik1990} employed a diffusion quantum Monte-Carlo (DQMC) approach.
Sadlej and coworkers are the only authors who investigated static dipole polarizabilities for other species than LiH using many-body perturbation theory (MBPT) and coupled-cluster theories combined with finite field  methods \cite{sadlej1991,diercksen1991}. Most of the above-quoted authors presented predictions obtained with different basis sets or at different levels of approximation, and we only report in Table \ref{tab:pola} values obtained with their most sophisticated model.

Values for the polarizabilities at $R_e$ are similar to those for $v=0$ for all species but CsH within 2\%, simply because the steepness of the polarizability functions around $R_e$ is compensated by the symmetric character of the $v=0$ wavefunction. For CsH, the 9\% difference comes from our equilibrium distance, which is slightly smaller than the experimental one. For LiH and NaH our predictions agree well with other values with the exception of the values of Gready {\it et al.} \cite{gready1977b} obtained with a too small basis set. The relatively wide variation of the theoretical values clearly emphasizes the high sensitivity of the polarizability to the size of the basis sets and to correlation effects. Our predictions of $\alpha_{\parallel}$ for heavier systems lie between the two quoted values of ref.\cite{sadlej1991}, while our values for $\alpha_{\perp}$ are slightly smaller. Note that the dipole moments calculated with their approximation (b) in that paper are closer to our values than their (a) values.

We are not aware of any polarizability measurement in alkali hydrides. An old experimental value of the polarizability anisotropy for LiH with a large error bar ($\gamma = 1.7 \pm 4.$ au) has been quoted  by Stevens and Lipscomb \cite{stevens1964} as an unpublished result from Klemperer {\it et al.}.

\begin{table}[h]
\begin{tabular}{|c|c|c|c|c|c|c|} \hline
    &                                          &$\alpha_{zz}$&$\alpha_{xx}$&$\alpha$&$\gamma$&$\mu_z$ \\ \hline
LiH &This work, for $v=0$                      &26.8&30.3&29.1&-3.5&5.87\\
    &This work                                 &26.0&29.8&28.6&-3.8&5.82\\
    &Gready {\it et al.} \cite{gready1977b}      &34.1&32.6&33.1&1.5&5.90\\
    &Lazzretti and Zanasi \cite{lazzeretti1981}&21.93&25.3&24.2&-3.4&6.0\\
    & Roos and Sadlej \cite{roos1982}          &26.3&29.3&28.3&-3.&5.90\\
    &Bishop and Lam \cite{bishop1985}          &26.3&29.7&28.6&-3.4&6.0\\
    &Vrbik {\it et al.} \cite{vrbik1990}        &24.6&30.9&28.8&-6.3&5.76\\
    &Maroulis and Bishop \cite{maroulis1986}   &21.8&25.1&24.&-3.3&6.0.\\
    &Sasagane {\it et al.} \cite{sasagane1990}      &22.9&29.4&27.3&-6.4&\\
    & R\'erat {\it et al.} \cite{rerat1992}     &26.9&30.8&29.5&-3.9&5.87 \\
    &Sadlej  and Urban (a) \cite{sadlej1991}   &26.7&30.&29&-3.2&5.909\\
    &Sadlej  and Urban (b) \cite{sadlej1991}   &26.81&30.01&28.94&-3.2&5.906\\
    &Wharton {\it et al.} \cite{wharton1962} (exp.)    && & & & 5.882 $\pm 0.003$ \\ \hline
NaH &This work, for $v=0$ &55.3&39.2&44.5&16.1&6.41\\
    &This work& 53.7&38.8&43.7&15&6.39\\
    &Sadlej  and Urban (a) \cite{sadlej1991}  &54.1&39.9&44.6&14.3&6.67\\
    &Sadlej  and Urban (b) \cite{sadlej1991}  &58.9&39.7&46.1&19.2&6.38 \\
    &Diercksen and Sadlej \cite{diercksen1991}&50.8&37.7&42.0&12.5&6.44\\
    &Dagdigian\cite{dagdigian1979} (exp.)&& & & & 6.4$\pm  0.07$\\ \hline
KH  &This work, for $v=0$ &72.0&44.6&53.7&27.4&8.11\\
    &This work  &71.8&44.5&53.6&27.3&8.14\\
    &Sadlej  and Urban (a) \cite{sadlej1991}&66.0&48.4&54.3&17.6&8.48\\
    &Sadlej  and Urban (b) \cite{sadlej1991}&85.0&50.5&62.0&34.5&8.08\\ \hline
RbH &This work, for $v=0$ &85.6&44.1&57.9&41.5&8.36\\
    &This work&85.&44.2&58.0&41.3&8.42\\
    &Sadlej  and Urban(a) \cite{sadlej1991}&71.4&48.2&56.0&23.1&9.04\\
    &Sadlej  and Urban (b) \cite{sadlej1991}&95.34&51.5&66.1&43.8&8.41\\ \hline
CsH &This work, for $v=0$ &91.6&34.7&53.7&56.9&8.22\\
    &This work&92.6&37.5&55.9&55.&8.67\\ \hline
\end{tabular}
\caption {Polarizabilities (in a.u.) and permanent dipole moments (in Debye) for the ground state of alkali hydrides, either computed for the $v=0$ level, or taken at the experimental equilibrium distance, compared to published theoretical and experimental data. Values of ref. \cite{gready1977b} corresponds to their "CI" calculation, in ref.\cite{sasagane1990} to their "MCTDHF[DQ]" calculations, in \cite{diercksen1991} to their "MBPT(4)" calculations, and in ref.\cite{sadlej1991} to their "T(CCSD)" (a) in the valence-shell approximation and (b) in a model including correlations from both valence and next-to valence shells.}
\label{tab:pola}
\end {table}

\section{Trends of the permanent and transition dipole moments of alkali hydrides}
\label{sec:dipole}

We report in Table \ref{tab:pola} our values for the permanent dipole moment $\mu_z$ at the experimental equilibrium distance of the ground state. We note that they are quite a bit larger than those of alkali dimers \cite{aymar2005} (ranging between 0.5 and 5~Debye), due to the large asymmetry of the charge distribution. Our values show good agreement with the limited experimental data available (currently only for LiH and NaH \cite{wharton1962,dagdigian1979}). Table \ref{tab:pola} reflects the situation which held until recently for the theoretical works: many of them were devoted to LiH, and only a few to heavier species.  In particular, we obtain a very good agreement with the most elaborate model of Sadlej and Urban \cite{sadlej1991}.

We also computed the $R$-variation of the permanent and transition dipole moments for a selection of electronic states. It is well-known that such functions represent sensitive tests for the accuracy of the computed electronic wave functions. As mentioned earlier, many theoretical works have been already published for these quantities, and we concentrate on systematic trends in their behavior along the series of alkali hydrides. It is worth noting that such an analysis can be performed due to the wealth of data concerning alkali hydrides, in contrast with the few data available for alkali dimers (see for instance ref.\cite{aymar2007}).

\begin{figure}
\includegraphics[width=0.8\columnwidth]{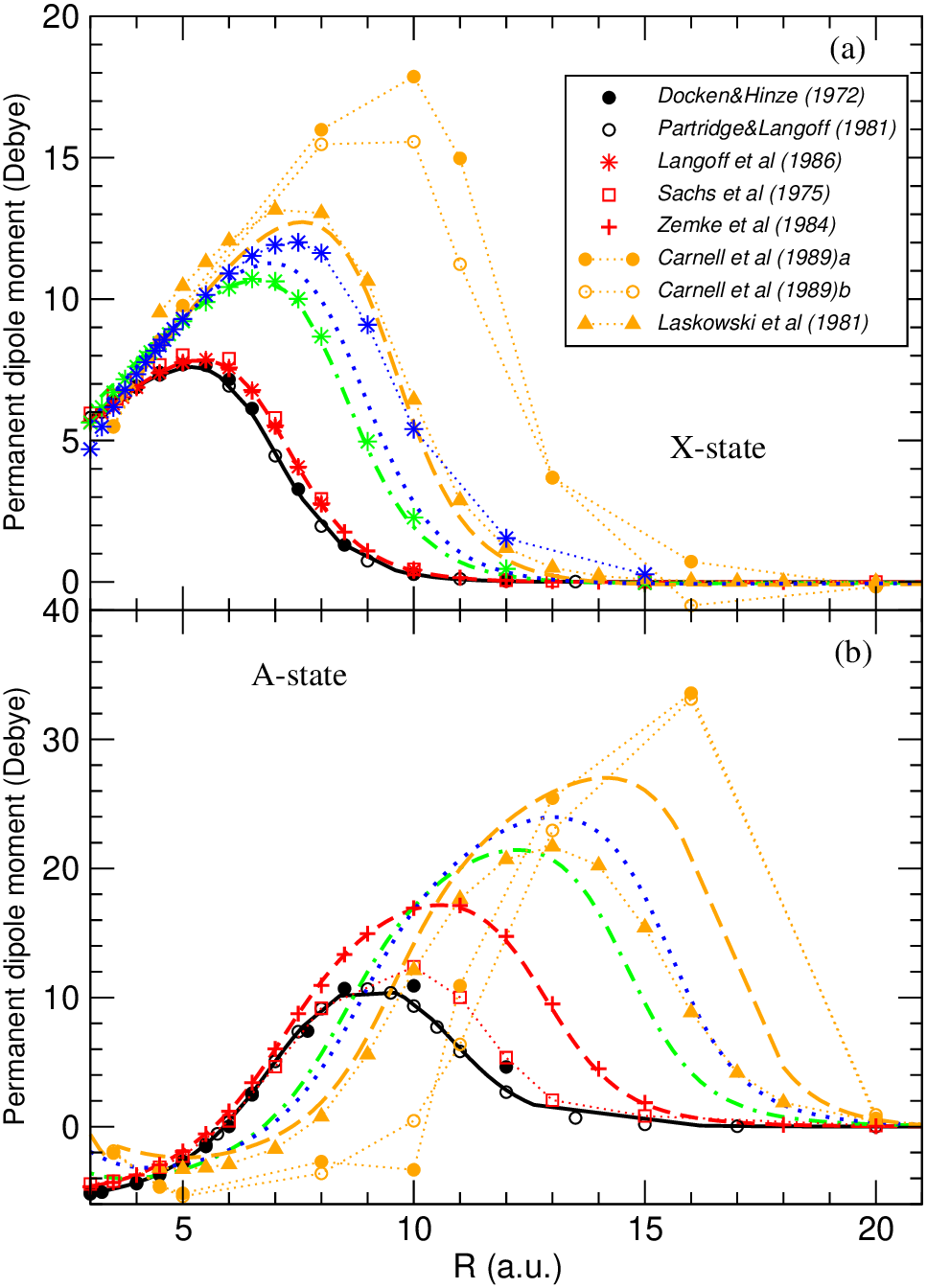}
\caption{\label{fig:permdipXA} Computed permanent dipole moments of $X^1\Sigma^+$ (upper panel) and $A^1\Sigma^+$ (lower panel) states as functions of the internuclear distance. Our results are: LiH (black full line), NaH (red dashed line),KH (green dot-dashed line), RbH (blue dotted line), CsH (orange long-dashed line).  Other determinations are displayed from ref.\cite{docken1972} (black full circles), ref. \cite{partridge1981} (black open circles), ref.\cite{sachs1975} (open squares), ref.\cite{zemke1984} (plus signs), ref.\cite{langhoff1986} (stars), ref.\cite{laskowski1981} (full triangles), and  ref.\cite{carnell1989} without or with relativistic effects (dotted line and full circles, or dotted line with open circles, respectively).}
\end{figure}

In the following figures, we display the computed $R$-variation of the permanent dipole moments for the $X$, $A$, $B$, $C$, $a$, and $b$ states of alkali hydrides, compared to other available studies. As expected the most studied state is the $X$ ground state (Fig. \ref{fig:permdipXA}(a)). Its permanent dipole moment shows similar regular variations with $R$ for all species, reaching a higher maximum amplitude at larger distance with increasing mass and polarizability. The agreement between all methods is generally satisfactory, with the exception of ref.\cite{carnell1989} where the maximum value is found to be about 20\% larger. The permanent dipole moments for the $A$ state (Fig. \ref{fig:permdipXA}(b)) vary like the one for the $X$ state, reaching their maximum value at larger distances. Several other determinations exist in the literature for the LiH and NaH $A$ state. We found a remarkable agreement for LiH with refs. \cite{docken1972,partridge1981}. In contrast, the maximum value found by Sachs {\it et al.} \cite{sachs1975} is almost two times smaller than ours. As for the ground state, the maximum value found by Carnell {\it et al.} for CsH is larger (by about 25\%) than ours, while the one by Laskowski {\it et al} \cite{laskowski1981} is about 25\% smaller than ours. To our knowledge, the only other calculations for KH, RbH, and CsH have been reported by Gad\'ea and coworkers in refs \cite{khelifi2002}, \cite{khelifi2002a}, and \cite{zrafi2006}, respectively. No tables of numerical data are available in these papers, but a satisfactory agreement is visible, as long as the eye can judge from their figures (see the discussion for RbH later in this section). Finally our calculations for both the $X$ and $A$ states of NaH are in excellent agreement with those of Zemke {\it et al.} \cite{zemke1984} who extracted the dipole moment functions from an analysis combining ab-initio and experimental results.

\begin{figure}
\includegraphics[width=0.8\columnwidth]{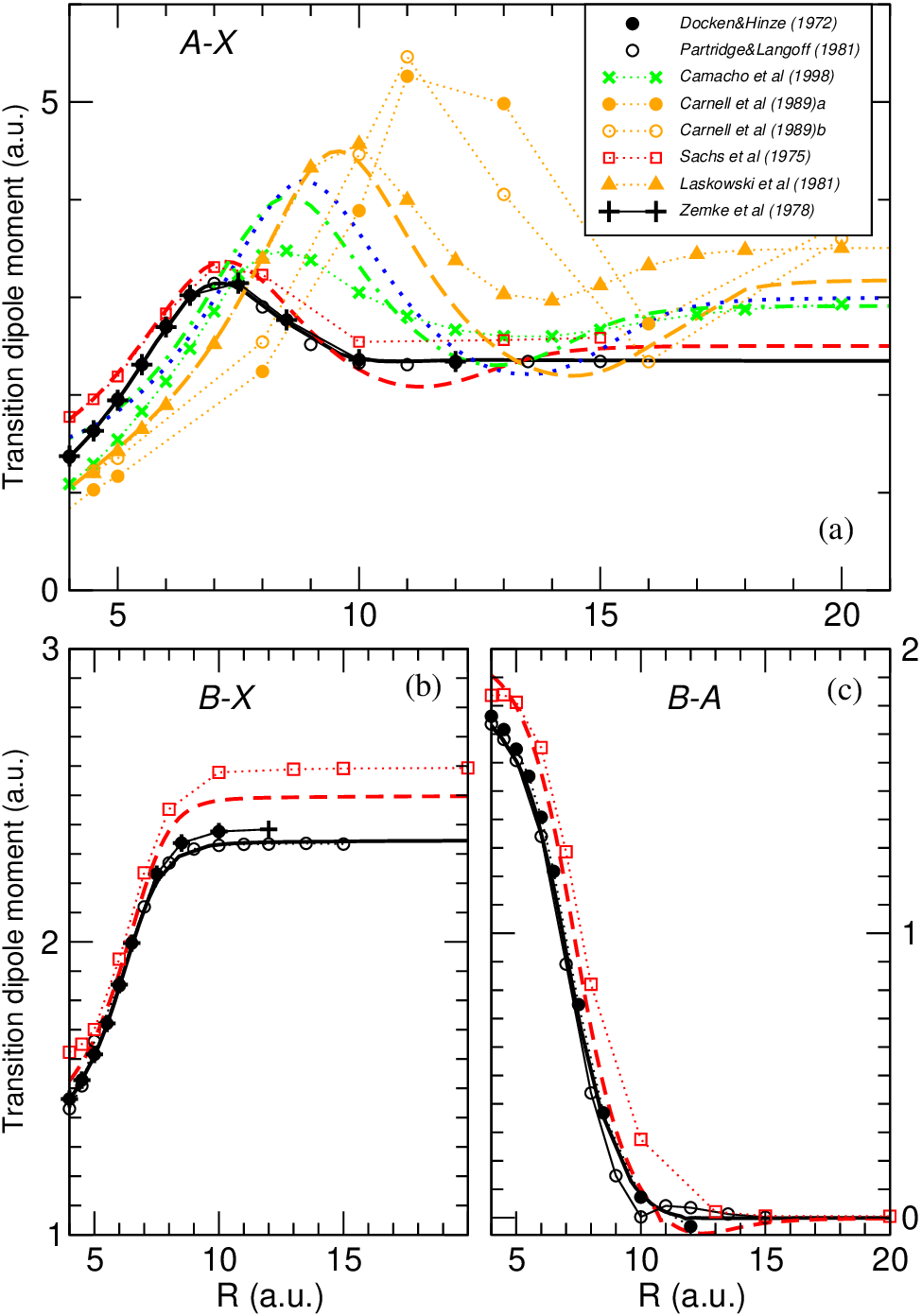}
\caption{\label{fig:transdipXAB} Computed transition dipole moments for (a) the $A^1\Sigma^+ - X^1\Sigma^+$ transition, (b) the $B^1\Pi - X^1\Sigma^+$ transition, (c) the $B^1\Pi - A^1\Sigma^+$ transition, as functions of the internuclear distance. Our results are: LiH (black full line), NaH (red dashed line),KH (green dot-dashed line), RbH (blue dotted line), CsH (orange long-dashed line). Other determinations are displayed from ref.\cite{docken1972} (black full circles), ref. \cite{partridge1981} (black open circles), ref.\cite{sachs1975} (open squares), ref.\cite{zemke1978a,zemke1978b} (plus signs), ref.\cite{camacho1998} (crosses), ref.\cite{laskowski1981} (full triangles), and  ref.\cite{carnell1989} without or with relativistic effects (dotted line and full circles, or dotted line with open circles, respectively).}
\end{figure}

As expected, the $R$-variation of the $A-X$ transition dipole moment (Figure \ref{fig:transdipXAB}a) reflects both variations of the $X$ and $A$ permanent dipole moments through the corresponding electronic wave  functions: two extremum are visible, whose positions correspond to those obtained for the $X$ and $A$ permanent dipole moment functions. The same figure also displays results concerning the transitions $B-X$ and $B-A$ involving the lowest $B^1\Pi$ excited state (Figure \ref{fig:transdipXAB}b,c).  An excellent agreement is found again with refs.\cite{docken1972,partridge1981} for all these transitions in LiH. Quite unexpectedly, the agreement with the values of Sachs {\it et al} \cite{sachs1975} for NaH looks better than for the $A$ permanent dipole moment. For CsH, the curve by Laskowski {\it et al} \cite{laskowski1981} is close to ours in the region of the maximum value of the $X$ dipole moment (below 10 a.u.), and not in the region of the one of the $A$ dipole moment (around 15 a.u.). As in the previous figure, results from ref.\cite{carnell1989} disagree with ours. Again, the agreement for the $A-X$ transition dipole moments seem satisfactory when looking at figures from refs.\cite{khelifi2002,khelifi2002a,zrafi2006} for KH, RbH, and CsH. While not displayed here, results are also available for transition dipole moments among the lowest triplet $\Sigma^+$ and $\Pi$ states ($c-a$, $b-a$, $c-b$) in LiH \cite{docken1972} and NaH \cite{sachs1975}, in good agreement with the present results. Experimental values are only available for $A-X$ and $B-X$ transitions in LiH \cite{zemke1978a,zemke1978b}, and our values are in excellent agreement with them. A determination of the $A-X$ transition dipole moment variation in KH has been recently extracted from the experiment by Camacho {\it et al.} \cite{camacho1998} (which is to our knowledge the only experimental data for this molecule) showing a reasonable agreement with our value.

\begin{figure}
\includegraphics[width=0.8\columnwidth]{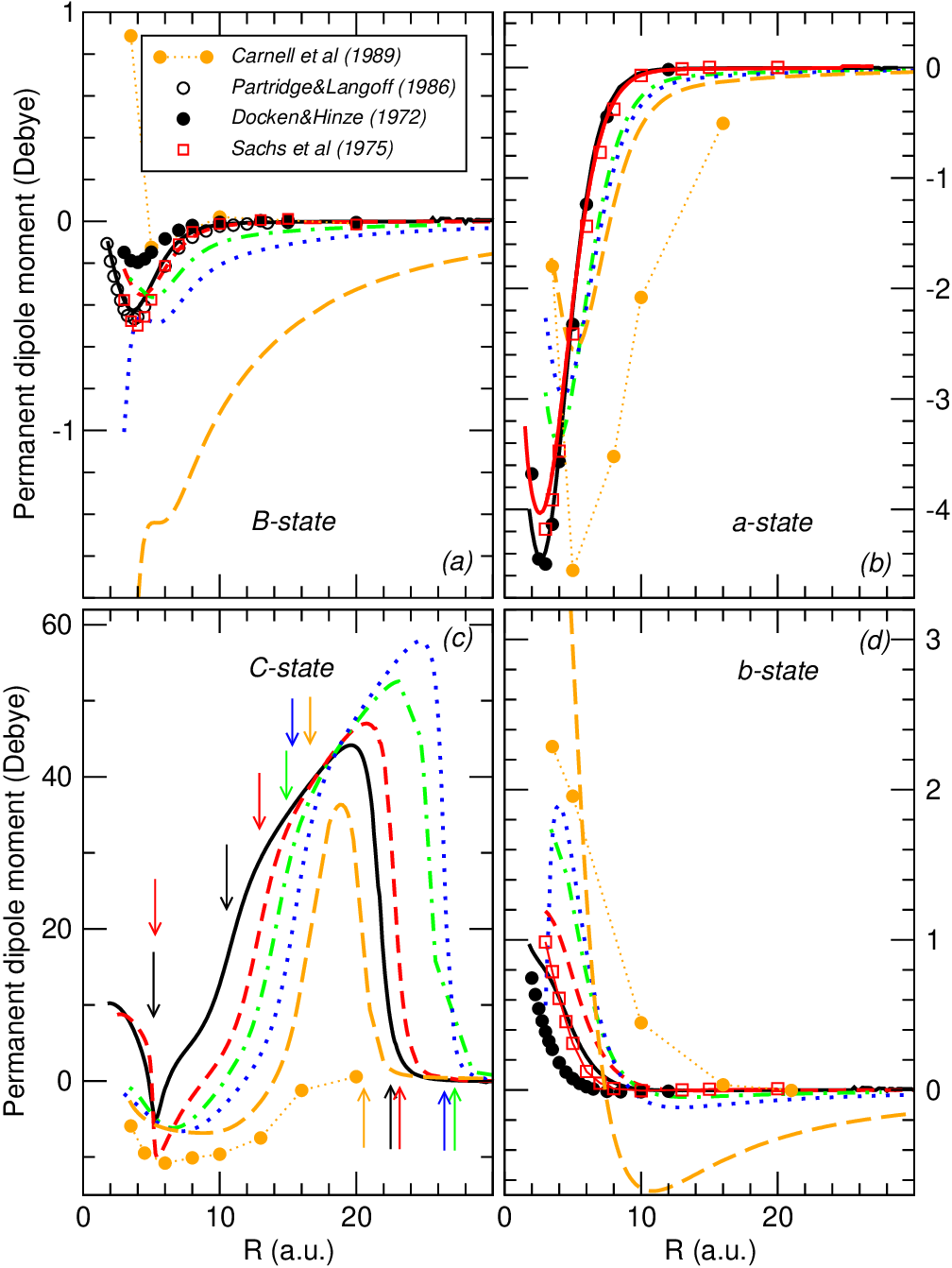}
\caption{\label{fig:permdipBCab} Computed permanent dipole moments of $B^1\Pi$, $C^1\Sigma^+$, $a^3\Sigma^+$, and $b^3\Pi$ states as functions of the internuclear distance. Our results are: Our results are: LiH (black full line), NaH (red dashed line),KH (green dot-dashed line), RbH (blue dotted line), CsH (orange long-dashed line).  Other determinations are extracted from ref.\cite{docken1972} (black full circles), ref. \cite{partridge1981} (black open circles), ref.\cite{sachs1975} (open squares), and  ref.\cite{carnell1989} without relativistic effects (dotted line and full circles). Arrows for each species correspond to the location of avoided crossings in the related potential curves in Figure \ref{fig:ionpair}.}
\end{figure}

In Figure \ref{fig:permdipBCab} we display the permanent electric dipole moment for higher excited singlet states ($B$ and $C$), and for the two lowest triplet $a$ and $b$ states. While less data are available for comparison, we observe the same trend than in the previous figures: the systematic investigation of Carnell {\it et al.} ref.\cite{carnell1989} really seems to miss most of the character of the excited electronic wave functions of CsH, while our results are in agreement with those of ref.\cite{partridge1981} for the $B$ state of LiH, and of ref.\cite{sachs1975} for the $B$, $a$, and $b$ states in NaH. A visual satisfactory agreement is also observed with the figures displayed in refs.\cite{khelifi2002,khelifi2002a,zrafi2006}.

\begin{figure}
\includegraphics[width=0.8\columnwidth]{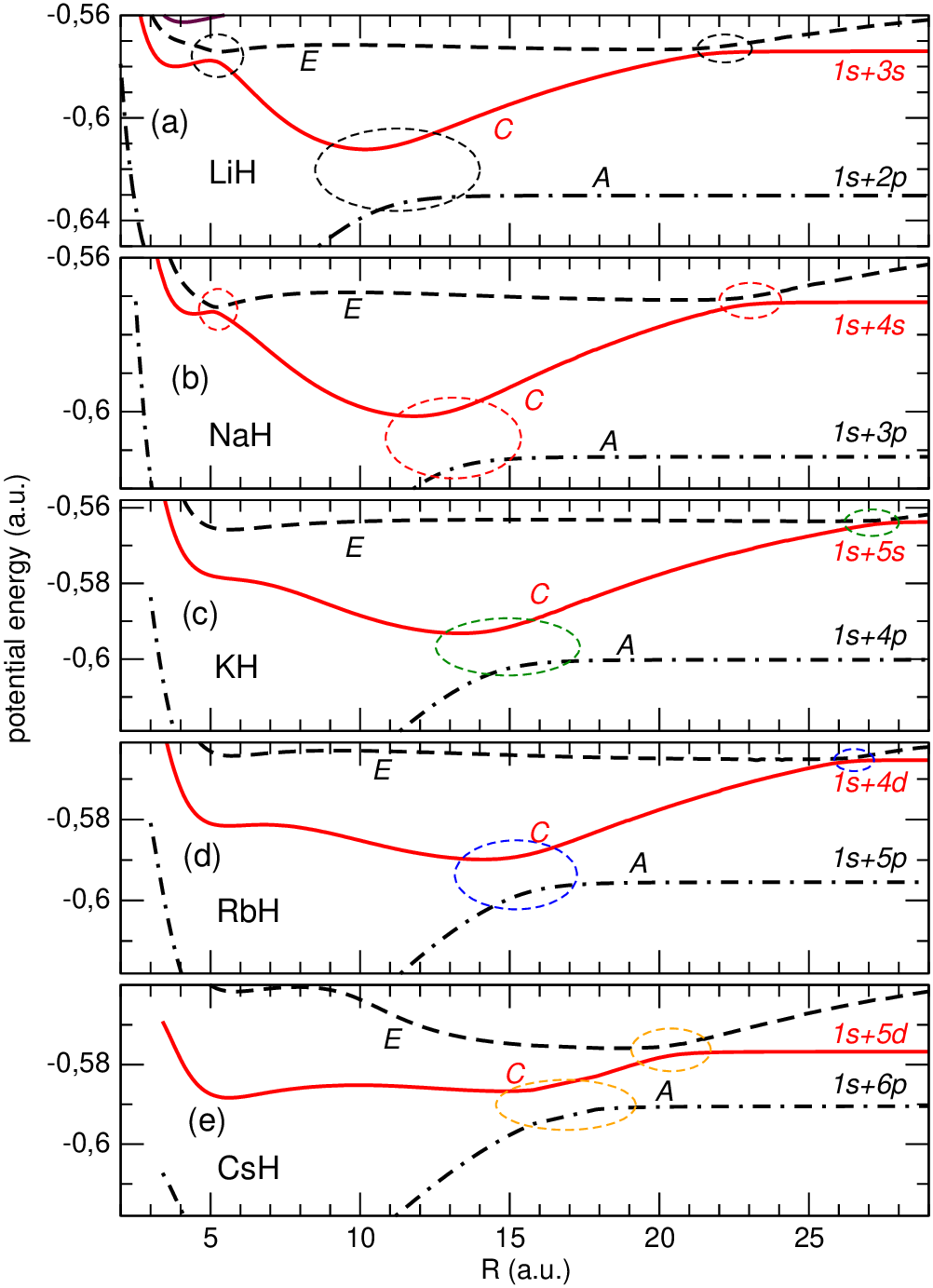}
\caption{\label{fig:ionpair} Potential curves of the alkali hydrides $^1\Sigma^+$ states in the region where the ion-pair contribution dominates the electronic wave function. The $A$, $C$, and $E$ states are drawn with dot-dashed, full, and dashed lines respectively. The origin of energies is taken at the double ionization limit of the atoms at infinite distance. Avoided crossings marked with circles correspond to the features discussed in Figure \ref{fig:permdipBCab}c and in Figure \ref{fig:transdipACE}.}
\end{figure}

As expected, all of these dipole moments vanish with increasing distance, with a smaller rate with increasing alkali mass. The $C$ state is the most excited state displayed in the figure - correlated to the second excited $1s+3s$, $1s+4s$, $1s+5s$, $1s+4d$, and $1s+5d$ dissociation limits for LiH, NaH, KH, RbH, and CsH, respectively- so its magnitude reaches larger values than for the other states. For all alkali hydrides, the $C$ dipole moment exhibits irregular variations, which is the well-known manifestation of the contribution of the ion-pair configuration in the electronic wave function (see Figure \ref{fig:ionpair}): as long as this ion-pair character dominates the wave function (between the avoided crossings marked with circles in Figure \ref{fig:ionpair}), the permanent dipole moment can take large values, until the distance where the covalent character shows up again. In LiH and NaH, the presence of the avoided crossing $C$ and $E$ states at short distances is also visible in Figure \ref{fig:permdipBCab}c (see arrows). Quite unexpectedly, the CsH dipole moment is not the largest one of the series, as the ion-pair character cannot develop towards distances larger than 22~a.u. due to the somewhat smaller energy spacing between the excited dissociation limits, compared to the other species.

\begin{figure}
\includegraphics[width=0.8\columnwidth]{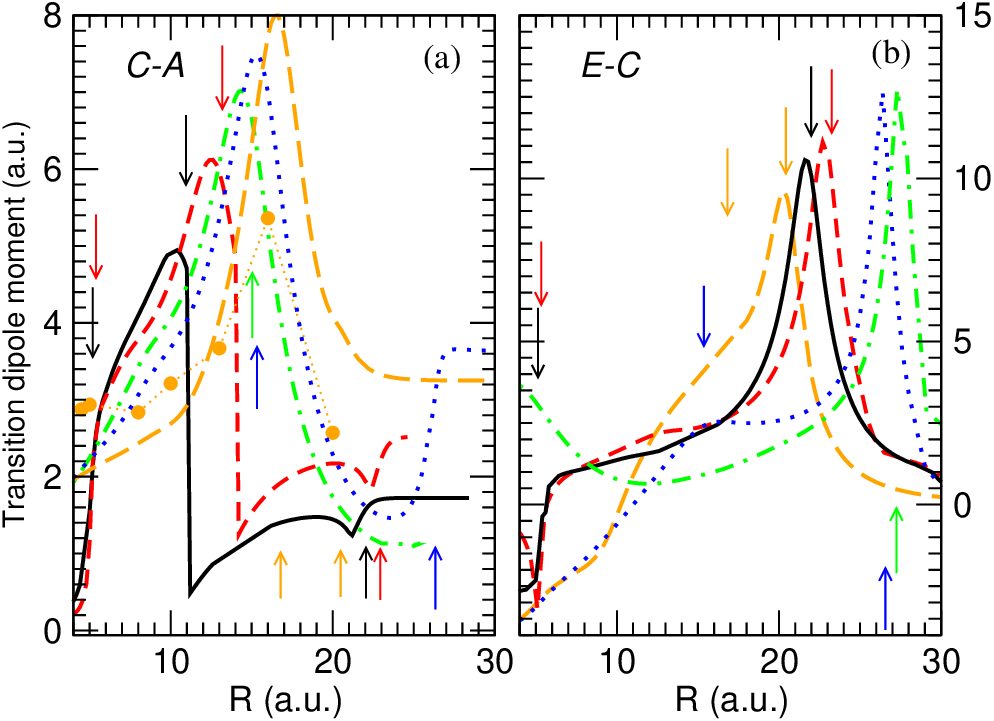}
\caption{\label{fig:transdipACE} Computed transition dipole moments for (a) the $C^1\Sigma^+ - A^1\Sigma^+$ transition, (b) the $E^1\Sigma^+ - C^1\Sigma^+$ transition as functions of the internuclear distance. Our results are: Our results are: LiH (black full line), NaH (red dashed line),KH (green dot-dashed line), RbH (blue dotted line), CsH (orange long-dashed line).  We reported in panel (a) the value from ref.\cite{carnell1989} without relativistic effects (dotted line and full circles). Arrows for each species correspond to the location of avoided crossings in the related potential curves in Figure \ref{fig:ionpair}.}
\end{figure}

The influence of the ion-pair character is also visible in the transition dipole moments among these excited electronic states, as visible on Figure \ref{fig:transdipACE} for the $C-A$ and $E-C$ transitions. The transition dipole moment functions change very abruptly with the internuclear distance (see arrows), which is the expected manifestation of local change of the electronic wave functions at the location of avoided crossings between the related potential curves.

The treatment of the ion-pair contribution is actually a sensitive issue for discussing the quality of the representation of the electronic wave functions of the molecules in various approaches. Indeed, such a character is not explicitly introduced in the standard Gaussian basis sets for neutral atoms, and would need in principle an infinite number of basis functions to be accurately represented. In the present work, we set up a large basis for the hydrogen atom, in order to reproduce the atomic energy levels, and hopefully to take in account as much as possible of the ion-pair character. An indication for this is our computed value of the hydrogen electronic affinity $EA=5938$~cm$^{-1}$, compared to the experimental value $EA^{exp}=6083$~cm$^{-1}$ \cite{pekeris1962}. Let us note that in ref.\cite{lee2000a} mentioned in Table \ref{tab:spectro} the authors used a $7s4p2d [6s4p2d]$ Gaussian basis set \cite{yiannopoulou1999} for the H atom, yielding a good determination of the hydrogen affinity  with deviation from the experimental \cite{pekeris1962} of $\sim 145$cm$^{-1}$ ,i.e. similar to our value, while the $1s$ and $2s$ states are less accurate than in our work, with $\Delta \sim  64$~cm$^{-1}$ and $\Delta E\sim 77$~cm$^{-1}$ respectively.

\begin{figure}
\includegraphics[width=0.8\columnwidth]{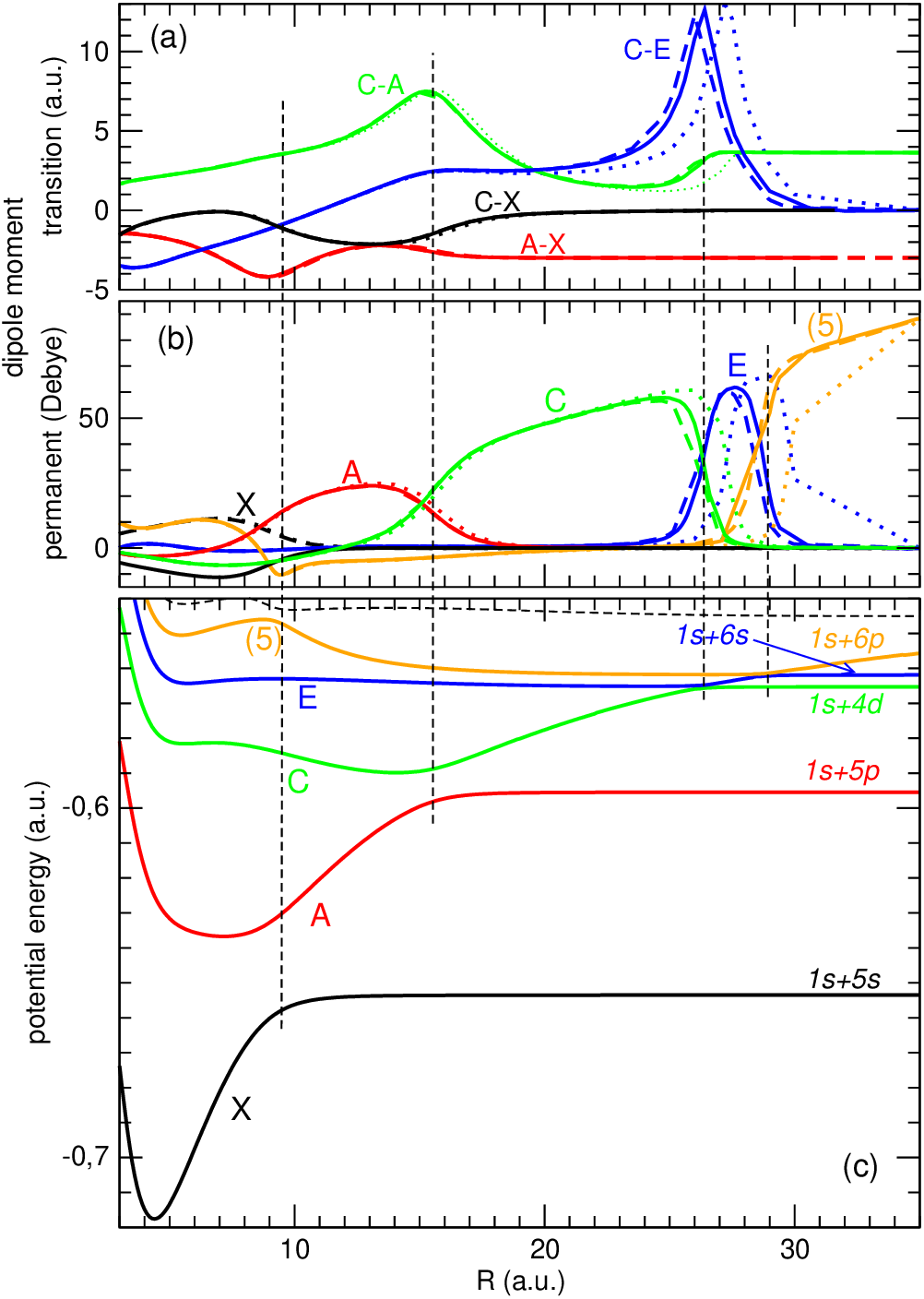}
\caption{\label{fig:permdip_rbh} Transition dipole moments (upper panel) and permanent dipole moments (middle panel) for the five lowest $^1\Sigma^+$ states of RbH computed in the present work (full lines), and by  Khelifi {\it et al.} \cite{khelifi2002a} without (with) empirical correction for hydrogen electron affinity (dashed and dotted lines, respectively). Related potential curves computed in the present work are displayed in the lower panel. The link between the presence of avoided crossings and the abrupt variations of the dipole moment is highlighted with vertical dashed lines.}
\end{figure}

The value of $EA$ determines the location of avoided crossings between potential curves with covalent and ion-pair character. In a series of papers, Khelifi {\it et al.} \cite{khelifi2002,khelifi2002a} and Zrafi {\it et al.} \cite{zrafi2006} addressed this issue in detail in KH, RbH, and CsH, respectively. We note that the overall agreement with the numerical values for RbH from Khelifi \cite{khelifi_pc} is excellent (Figure \ref{fig:permdip_rbh}), which is not surprising as our methods are quite close. Using a diabatization procedure, they were able to build an effective Hamiltonian which  separates the covalent and the ion-pair characters of the electronic states. Then they account for their discrepancy on the hydrogen $EA$ (405~cm$^{-1}$ in their case) as an empirical negative correction added to the corresponding diagonal element of the effective Hamiltonian. The related avoided crossings in the adiabatic potential curves are then shifted towards larger internuclear distances, as well as the associated abrupt variations in the transition dipole moment functions (see dashed lines in Figure \ref{fig:permdip_rbh} for the case of RbH). In contrast, we see from our calculations that the increase of the basis set size, which yields a value for the hydrogen $EA$ as close as 145~cm$^{-1}$ from the exact value, only provides a small shift of these patterns toward large distances compared to the results of ref.\cite{khelifi2002a} without the empirical correction. As tentative interpretation, we estimate that the position of the avoided crossing between the $E$ and the $C$ curves around 26.5$a_0$ (Figure \ref{fig:permdip_rbh}c) would be shifted only by about 0.4$a_0$ if we would lower our $E$ potential curve down by 145~cm$^{-1}$. However in our results, the associated peak in the $C-E$ transition dipole moment around 26.5$a_0$ (Figure \ref{fig:permdip_rbh}c) is separated by about 0.8$a_0$ from the one of ref.\cite{khelifi2002a} when the empirical correction is included. Therefore, it is not obvious if such a global energy shift of the ionic configuration accurately describes the ionic configuration as this correction does not change the electronic wave function of the negative ion, as also illustrated for instance by the excessive increase of the CsH potential well depth by 462~cm$^{-1}$.

\section{Conclusion}

In this paper, we computed the electronic properties of alkali hydrides from LiH to CsH, including potential curves, permanent and transition dipole moments, and static dipole polarizabilities. We performed a systematic investigation which allowed us to ensure a similar numerical accuracy for all species, in the same spirit as the numerous studies published by Gad\'ea and coworkers. Our results are found in good agreement with available experimental data, so that our systematic computations for all species are useful to estimate the accuracy of other available theoretical results. Apart for LiH, we determined the variation of the polarizabilities with internuclear distance for the first time, and their values for vibrational levels still await for experimental confirmation. Furthermore, general trends of these properties for the whole series of alkali hydrides have been demonstrated. Finally, we discussed the importance of the accuracy of the representation of the hydrogen electronic affinity in the long-range dynamics of the molecular excited states. In particular, we suggested that the empirical model based on a diabatization procedure by Gad\'ea and coworkers may not address this issue in the same way as our calculation using an extended basis set for the hydrogen atom. Recent developments may soon bring comparison between theory and experiment into reach. For instance, an intense supersonic beam of LiH molecules has been set up in the perspective of a Stark deceleration experiment \cite{tokunaga2007}. Also, very precise Stark spectroscopy has been demonstrated in an ensemble of ultracold KRb molecules in their ground state, yielding accurate values for permanent and transition dipole moments \cite{ni2008}.

\section*{acknowledgments}
The authors thank N. Khelifi for providing us with his data for RbH. This work is performed in the
framework of the network "Quantum Dipolar Molecular Gases" (QuDipMol) of the EUROCORES/EUROQUAM program of the European Science Foundation. J.D. acknowledges partial support of the French-German University (http://www.dfh-ufa.org).


\end{document}